# When Cooperation Should End: Maneuver Coordination Cancellation for Connected Automated Driving

Rafael Molina-Masegosa[a], Sergei S. Avedisov[b], Miguel Sepulcre[a],
Javier Gozalvez[a], Onur Altintas[b]
[a] Universidad Miguel Hernandez de Elche (UMH), Spain, {rafael.molinam, msepulcre, j.gozalvez}@umh.es
[b] Toyota Motor North America R&D – InfoTech Labs, Mountain View, CA, USA, {sergei.avedisov, onur.altintas}@toyota.com

***Abstract*—Maneuver coordination is essential for cooperative connected automated driving, enabling vehicles to negotiate maneuvers and interactions through V2X communication. While prior work has largely focused on how to initiate and execute coordinations, considerably less attention has been given to how ongoing coordinations should be terminated when they become unsuitable. This paper introduces the first complete design and implementation of maneuver coordination cancellation, including a state machine, message set, and decision-making logic. Our evaluation shows that cancellation significantly reduces the time vehicles spend in coordinations that cannot succeed, allowing them to become available for new maneuvers sooner. This increases the number of triggered coordinations and improves the number of successful maneuver coordinations. Overall, the study demonstrates that maneuver coordination cancellation improves cooperative driving, and establishes a foundation for further refinements that can enhance the efficiency and robustness of connected automated driving.**

## I. Introduction

Connected and Automated Vehicles (CAVs) are envisioned to play a central role in shaping future transportation systems. By leveraging vehicle-to-everything (V2X) communications, vehicles can coordinate their maneuvers, such as lane changes or merging, with the aim of reducing conflicts, improving traffic efficiency and enhancing safety. Maneuver coordination allows vehicles to account for the intentions of others and coordinate their behavior accordingly, transforming individual driving decisions into cooperative interactions among multiple participants. Despite its potential, maneuver coordination faces multiple challenges. It requires real-time negotiation among vehicles, which could operate with incomplete and potentially outdated information, where timing, reliability of communication, and the dynamic nature of traffic all play critical roles.

Previous research has investigated how to design maneuver coordination mechanisms that ensure efficient and safe cooperation among vehicles. Contributions include the design of message formats [1], negotiation strategies [2], and coordination protocols [3] among others. Standardization efforts in Europe [4] and the United States [5] have further advanced this field by defining reference use cases and communication requirements. Nevertheless, previous research concentrates on the initiation and execution of maneuvers, often trying to optimize how vehicles negotiate and commit to a coordinated maneuver. In contrast, considerably less attention has been given to mechanisms that manage the failure of a coordination process, or to strategies that ensure the best outcome in the presence of disturbances. The study presented in [6] identifies adaptability to unexpected traffic changes as one of the key challenges for effective maneuver coordination. Situations such as sudden variations in surrounding traffic can invalidate an ongoing coordination, making its safe completion uncertain. In these circumstances, the ability to cancel a coordination offers significant potential. An explicit cancellation mechanism could allow vehicles to disengage from maneuvers that have become unsuitable, preventing unnecessary delays and reducing inefficiencies. Furthermore, cancellation enables vehicles that were engaged in a coordination that would not succeed to become available earlier for other maneuvers, while also allowing nearby traffic to participate in alternative and potentially more beneficial coordinations. This additional flexibility enhances both the robustness and the overall efficiency of cooperative driving.

The SAE standard introduces the concept of cancellation and provides a high level description [5], opening the door for implementers to define when and how a cancellation is triggered. In this paper, we introduce the first complete and functional design for maneuver coordination cancellation. Our design includes a formal definition of the state machine, message set, and communication protocols required to effectively trigger a cancellation. In addition, we propose an initial decision-making mechanism for determining when a cancellation should be initiated. Finally, we evaluate the framework through system-level simulations covering a wide range of traffic densities, demonstrating its effectiveness in improving maneuver coordination and highlighting promising directions for future research and development.

## II. Maneuver coordination implementation

This section introduces a complete implementation of maneuver coordination that enables testing in realistic traffic scenarios and has been presented in our earlier work [6]. We describe the main elements that define the coordination process, including the state machine, the messages exchanged between vehicles, the roles assigned to participants, the possible outcomes of a coordination, the conditions that need to be satisfied to trigger a coordination, and how vehicles move during the execution of the maneuver. These aspects are outlined in general terms and then specified for the coordinated lane change maneuver in a multilane highway scenario, which serves as the reference use case in our evaluation.

The maneuver coordination framework is implemented as a state machine model, following SAE guidelines and extending the approach in [6]. The diagram in Fig. 1 summarizes the complete state machine with the states for each vehicle role, messages transmitted in each state, and state transition conditions. Vehicles operate primarily in the Intent Sharing state, where they are not engaged in coordination but periodically broadcast Intent messages describing their intended actions for the next few seconds. These messages, generated according to predefined rules [7] with a maximum period of 1 s between messages, enable nearby vehicles to assess the traffic context and decide whether a coordination should be initiated.

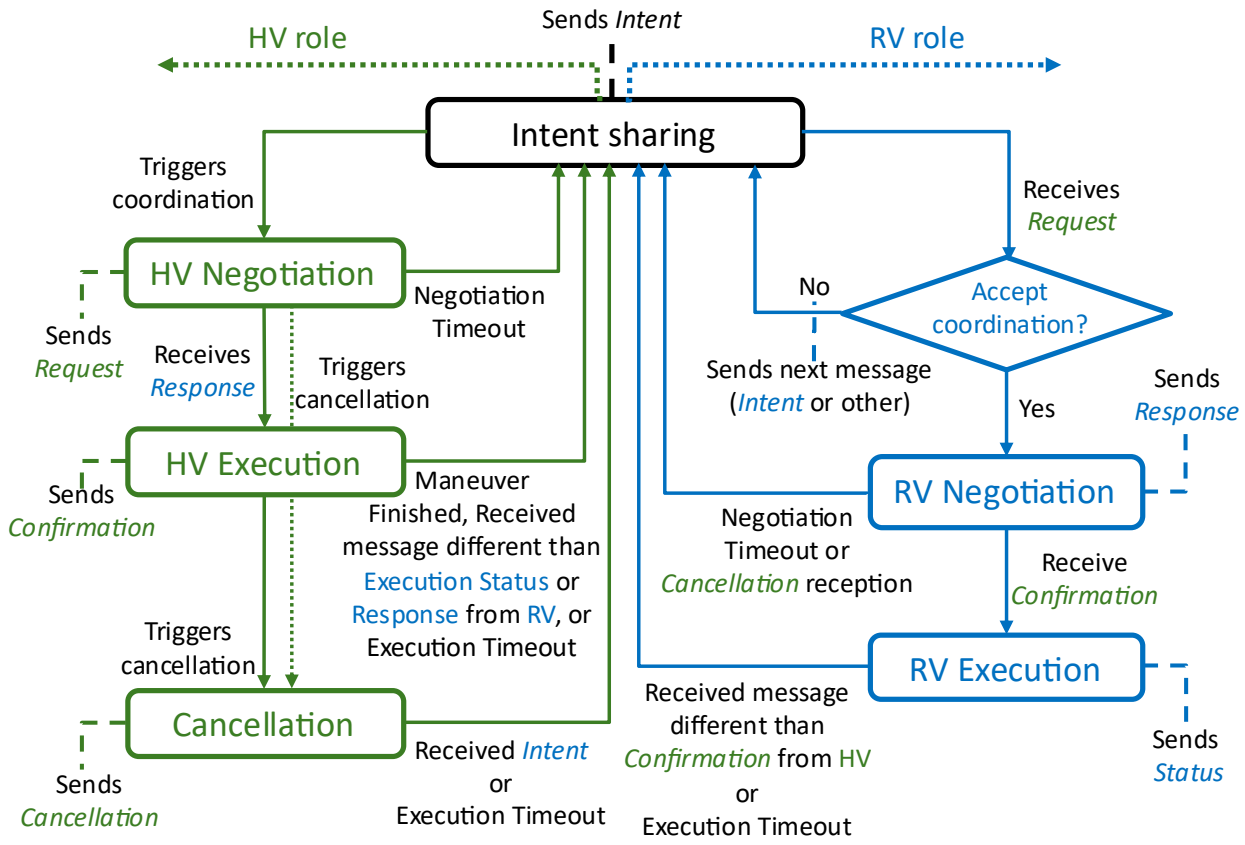

Fig. 1. Vehicle roles, state machine and messages for maneuver coordination implementation. Cancellation state, message and transitions have been added for the cancellation implementation in this work.

When a vehicle initiates coordination, it assumes the role of the Host Vehicle (HV) and transitions to the HV Negotiation state. In this state, the HV transmits Request messages every 100 ms (the minimum transmission period) until one of two outcomes occurs: (1) the designated Remote Vehicle (RV) accepts the request, or (2) the Negotiation Timeout expires, measured from the Coordination Triggering (CT) time. If the timeout elapses without a response, the HV aborts the attempt and reverts to the Intent Sharing state. If the RV accepts, the HV enters the HV Execution state, where it broadcasts Confirmation messages every 100 ms. This continues until (1) the maneuver is completed, (2) a message is received from the RV that is neither a Response nor an Execution Status message, or (3) the Execution Timeout is reached. The Execution Timeout is derived from the Coordination Intended Finish (CIF) time plus a predefined margin to account for execution delays. At this point, the HV returns to Intent Sharing. To ensure synchronization and reliability, all Request, Response, Confirmation, and Execution Status messages include the identifiers of the HV and RV, the maneuver ID, and the CT and CIF times. Furthermore, whenever a vehicle receives any of these messages, it must transmit its next message within the minimum 100 ms interval, regardless of whether it is an Intent or coordination message.

When a vehicle in the Intent Sharing state receives a Request, it must decide whether to act as the RV. If it declines, it remains in Intent Sharing and resumes Intent broadcasts within the 100 ms minimum interval, signaling its rejection. If it accepts, it transitions to the RV Negotiation state and begins transmitting Response messages every 100 ms until either (1) a Confirmation from the HV finalizes the negotiation, or (2) the Negotiation Timeout expires, in which case the vehicle returns to Intent Sharing. Upon receiving Confirmation, the RV transitions to the RV Execution state, where it executes the maneuver and transmits Execution Status messages every 100 ms. Execution continues until either the HV exits HV Execution (signaled by a message other than Confirmation) or the Execution Timeout is reached, after which the RV reverts to Intent Sharing.

The maneuver coordination process may result in three possible outcomes. Unsuccessful negotiation occurs when the HV and RV fail to complete the negotiation within the allowed time, causing both vehicles to return to the Intent Sharing state. Unsuccessful execution arises when a previously agreed maneuver cannot be carried out to completion, either due to timeout expiration or premature termination, after which both vehicles also revert to Intent Sharing. We group these two outcomes under unsuccessful coordination. Since unsuccessful negotiations have a negligible impact on the evaluations presented in this work, the reader may consider all instances of unsuccessful coordination as unsuccessful executions. Finally, successful coordination is achieved when the maneuver is executed as planned within the allowed time, with both vehicles seamlessly transitioning back to Intent Sharing once completed.

Before initiating a maneuver coordination, a vehicle must determine whether doing so is both feasible and beneficial. This requires anticipating how the traffic situation will evolve in the short term and assessing the planned maneuver against the predicted conditions. The goal is to maximize the chances of initiating coordinations that can be successfully completed, while reducing the occurrence of unnecessary or unfeasible attempts. A key step in this process is future traffic prediction, which estimates the likely trajectories of the ego vehicle (as a potential HV) and surrounding vehicles over the next few seconds, including their positions, speeds, and accelerations. Although achieving high accuracy is challenging, improved prediction directly enhances coordination efficiency. In our implementation, we assume an upper-bound prediction following [8] to focus on the design of the maneuver coordination process. The behavior of a vehicle can change at any time, and even minor deviations from the predicted trajectories, such as a vehicle accelerating earlier than anticipated or a lane becoming unexpectedly occupied, may lead to Unsuccessful executions. It is then necessary to introduce cancellation mechanisms so that vehicles can abort maneuver coordinations when they estimate they would be unsuccessful due to unexpected changes in traffic conditions that affect the maneuver coordination.

After predicting the trajectory of nearby vehicles, a vehicle must assess whether initiating a maneuver coordination is both beneficial and feasible. The aim is to ensure that coordination is triggered only when cooperation is necessary and when the maneuver contributes positively to overall traffic flow. Since a coordination may last several seconds, this evaluation considers effects over a time horizon rather than relying on instantaneous conditions. Treiber shows in [9] that traffic flow is optimized when vehicles choose lanes with speeds that best match their desired speeds.

In the coordinated lane change maneuver analyzed in this study, the incentive criterion for vehicles to change lanes is to move to a lane whose speed better matches their desired speed. Vehicles use V2X information to anticipate traffic conditions and estimate lane speeds. A comfort criterion is also defined, allowing a lane change only when the resulting deceleration imposed on surrounding vehicles is low enough to avoid inefficient disturbances. In this context, a coordinated lane change is considered beneficial when the ego vehicle intends to change lanes according to the incentive criterion but cannot safely complete the maneuver without cooperation or without causing disturbances to surrounding traffic, as defined by the comfort criterion.

If a vehicle estimates that the maneuver coordination is beneficial under the predicted traffic conditions, it then evaluates whether it is feasible. To do so, the vehicle evaluates whether the lane change can be executed using the predicted trajectories. In this study, we consider that vehicles can execute a lane change if the potential acceleration of vehicles involved in the maneuver can be handled without a harsh braking even if such braking does not pose a safety risk as it could disrupt traffic flow. The vehicles that may need to decelerate following a lane change are the ego vehicle $V_{ego}$ and the vehicle $V_{bn}$ immediately behind $V_{ego}$ in the adjacent (new) lane. A lane change is then considered feasible if $\tilde{a}_{ego}$, $\tilde{a}_{bn} > b_{comf}$, where $\tilde{a}_{ego}$ and $\tilde{a}_{bn}$ represent the accelerations that $V_{ego}$ and $V_{bn}$ would experience after the lane change under consideration. $b_{comf}$ is set to -3 m/s$^2$ in this study. The feasibility assessement also includes checking that neither the HV nor the RV is already engaged in another coordination process, and that sufficient time is available to complete both the negotiation and execution phases. If both benefit and feasibility assessments are fulfilled, the vehicle initiates the maneuver coordination process presented in section II.

Another important aspect of a maneuver is how vehicles move during the execution of the maneuver, which must be specified in advance to ensure safety and efficiency. In a coordinated lane change, the HV is the vehicle intending to change lanes, while the RV is the vehicle immediately behind in the target lane. After negotiation has been completed, the RV initiates a controlled deceleration for a limited time, after which it continues at a constant speed until the maneuver is finished. The HV only executes the lane change once sufficient gap has been created by the RV and when the maneuver is judged to be both safe and beneficial, preserving both safety and efficiency in the overall process.

## III. DESIGN OF MANEUVER COORDINATION CANCELLATION

To evaluate the potential of maneuver coordination cancellation, it is essential to develop a complete and functional design and implementation. Existing standards have introduced the concept of cancellation and provided initial descriptions, but these remain at a high level [5], opening the door for implementers to define when and how a cancellation is triggered. In this work, we define the elements that a complete implementation must include. Fig. 2 illustrates a modular architecture of these elements, organized into two main components: the cancellation decision-making module and the cancellation management module. The cancellation decision-making module is responsible for monitoring an ongoing coordination and determining when the cancellation process should be initiated. This decision is based on a set of conditions that leverage future traffic predictions, an existing function already used in maneuver coordination to detect the need for coordinations and to support intent sharing. Following the feasibility criterion previously defined, the module assesses whether the current coordination remains feasible or if a cancellation must be triggered. Once a cancellation has been triggered, the cancellation management module governs the subsequent actions that are next presented. In addition, the module defines how vehicles move while the cancellation is being completed, guaranteeing a safe and efficient coordinated disengagement from the maneuver.

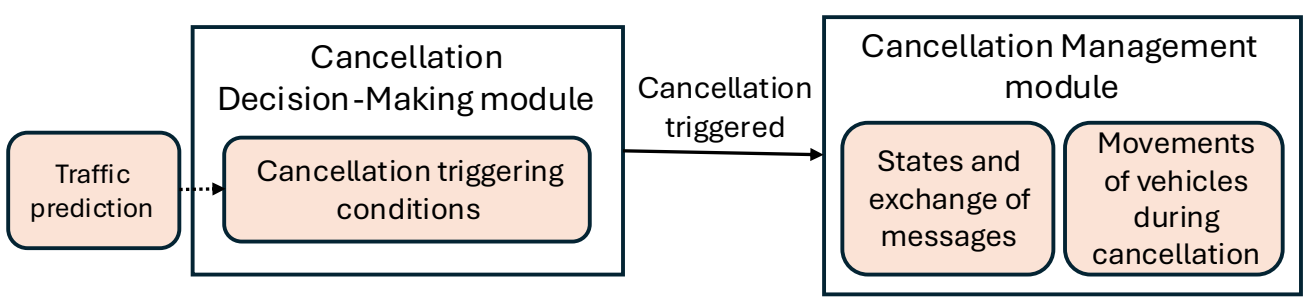

Fig. 2. Modular architecture of the complete maneuver coordination cancellation implementation

### A. Cancellation management module

The cancellation management module governs the protocol and state machine operations required once a cancellation has been triggered. To support its implementation, we extend the maneuver coordination state machine with the new elements needed for cancellation, as shown in Fig.1. On the HV side, the new Cancellation state is reached whenever the Cancellation Decision-Making module decides to interrupt an ongoing coordination, regardless of whether the HV is currently in the HV Negotiation or HV Execution state. Once in the Cancellation state, the HV begins transmitting Cancellation messages with the minimum period of 100 ms. This transmission continues until either the RV replies with a new Intent message or the Execution Timeout is reached, after which the HV returns to Intent Sharing. On the RV side, reception of a Cancellation message from the HV forces an immediate abortion of the coordination, whether the RV is in RV Negotiation or RV Execution. The RV then reverts to the Intent Sharing state. To guarantee timely confirmation, the RV is required to generate and transmit its next message within 100 ms of receiving a Cancellation, even if the next scheduled message according to intent-sharing rules [7] would normally be delayed for up to one second. This ensures that the HV is promptly informed that the cancellation has been completed at the RV's side. Accordingly we add a new possible outcome to the maneuver coordination described in Section II: coordination cancelled. This outcome is considered separately from those previously defined in Section II, which are successful coordination and unsuccessful coordination. During the coordinated lane change analyzed in this study, the HV does not modify its planned mobility while the RV aborts the planned deceleration to create a gap in the target lane upon receiving a Cancellation message. This is done to guarantee both safety and efficiency, allowing vehicles to disengage smoothly from an interrupted coordination.

### B. Cancellation decision-making module

The Cancellation Decision-Making module must detect when an ongoing maneuver coordination can no longer be successfully completed due to unexpected traffic changes. Its design must address two objectives: detecting as early as possible those coordinations that are likely to fail, and avoiding the cancellation of maneuvers that could still succeed. However, this module relies on the traffic prediction introduced in Section II. Because traffic prediction is inherently imperfect, these objectives cannot always be fulfilled with perfect accuracy, as consecutive and uncontrollable changes in traffic may render a maneuver unfeasible at one moment but feasible again shortly thereafter. Given this complexity, our initial implementation adopts a pragmatic and well-justified approach to coordinated lane changes that favors simplicity and reliability without sacrificing effectiveness. Specifically, a cancellation is triggered whenever the HV no longer has the incentive to change lanes. This criterion directly ties cancellation decisions to the same underlying logic that guides maneuver initiation, ensuring coherence between both processes.

It is important to note that cancellations are inherently associated with unstable situations. A maneuver coordination typically lasts only a few seconds, so if the need for cancellation arises after the maneuver was initially deemed feasible, this reflects a rapid change in incentives caused by evolving traffic ahead. In such contexts, the incentive to change lanes may disappear temporarily and reappear again shortly thereafter.

## IV. Simulation Framework

To evaluate the effectiveness of maneuver coordination cancellation, we run a comprehensive set of traffic simulations based on the full implementation described in Sections II and III. Experiments are performed on a custom simulation platform that couples a V2X network simulator with a vehicle mobility simulator [8]. V2X communication is simulated with the ns-3 network simulator, while vehicle dynamics are produced by a modification of the VANET Highway Mobility module for ns-3 [10]. The traffic scenario is a six-lane highway that is five kilometers long, with three lanes in each direction and periodic boundary conditions so that vehicles leaving one end of the road reappear at the other with the same state. We utilize several traffic densities from 10 to 40 vehicles per kilometer per lane. Traffic is composed of 80% passenger cars and 20% trucks. Desired speeds are sampled from a uniform distribution centered at 120 km/h for cars and 80 km/h for trucks, with a plus or minus 20% deviation. Each configuration is repeated for 20 independent runs of 200 seconds each to obtain statistically robust results.

Future traffic predictions are computed over a 5 second horizon and use context information from all vehicles located within 500 m of the ego vehicle. Because the prediction time window constrains the timing of a coordination, the Coordination Intended Finish time must lie within that horizon of 5 seconds. We adopt a fixed interval of 3 seconds between the Coordination Intended Finish and the Execution Timeout, so the total elapsed time from Coordination Triggering to maneuver completion can extend up to 8 seconds. The negotiation margin, defined as the interval between the Coordination Triggering time and the Negotiation Timeout, is set to 1 second. These timing choices follow previously validated configurations in [6]. During maneuver execution, the RV applies a controlled deceleration of 2 $m/s^2$ for 1 second and then maintains a constant speed until the coordination ends. To avoid interference between simultaneous coordinations, a minimum separation of 500 m is enforced between vehicles participating in different active coordinations. Thus, a vehicle will not start a coordination if another vehicle involved in an ongoing coordination is closer than this distance.

## V. Results and Discussion

This section presents the results obtained using the simulation framework introduced in Section IV. We first analyze how enabling cancellation affects the duration of coordinations that end unsuccessfully. The goal is to assess whether cancellation can reduce the time vehicles remain engaged in such unproductive interactions. Since cancellation does not prevent all unsuccessful coordinations, we consider all unsuccessful and cancelled coordinations jointly for a fair comparison between evaluations with and without cancellation. Fig.3 shows the average duration and $10^{th}/90^{th}$ percentiles of unsuccessful and cancelled coordinations for both settings. On average, cancellation reduces the duration of all unsuccessful and cancelled coordinations by half. This substantial reduction demonstrates the ability of cancellation to free vehicles from unproductive interactions more quickly, thereby improving system responsiveness and creating additional opportunities for new, potentially successful coordinations.

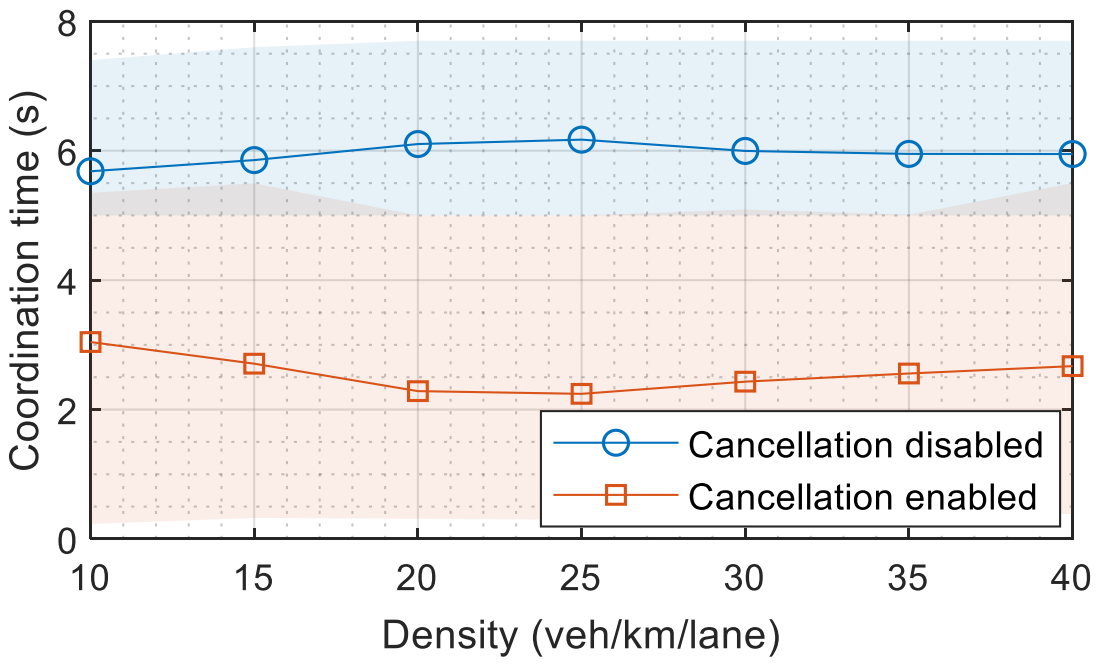

Fig. 3. Coordination time from triggering to abortion for all unsuccessful and cancelled coordinations. Solid lines indicate averages; shaded areas show 10th and 90th percentiles.

This capability is particularly valuable in situations where rapid maneuvering or enhanced responsiveness is critical. Moreover, earlier cancellations free capacity in the coordination process, leading to an overall increase in the number of triggered coordinations. Fig. 4 presents the number of triggered coordinations per vehicle per hour with cancellations disabled and enabled. We observe that, under medium and high traffic densities, the number of triggered coordinations increases by approximately 20-30% when cancellation is enabled. The increase is less pronounced in low-density scenarios, where vehicles naturally require fewer coordinations for lane changes. It is worth noting that the highest number of triggered coordinations per vehicle per hour in Fig. 4 occurs at medium traffic densities. At low densities, vehicles require fewer coordinations to change lanes due to the reduced congestion, while at high densities, there are fewer opportunities for coordinations to be feasible with a controlled deceleration by the RV. Maneuver coordination cancellation can also improve the total number of successful coordinations. Fig. 5 reports the number of successful, unsuccessful, and cancelled coordinations with cancellation enabled and disabled. Enabling cancellation leads to a 5-7% rise in successful coordinations for densities between 20 and 35 vehicles/km/lane, an improvement that confirms the added value of the proposed cancellation. The number of unsuccessful coordinations that are not cancelled is reduced to about one third, due to the early cancellation of most of them. However, we also observe a high number of cancelled coordinations when cancellation is enabled, even higher than the number of unsuccessful coordinations without cancellation. This indicates that many of the extra coordinations triggered thanks to the additional time gained by early cancellations do not end successfully and are likely to be cancelled as well. This shows there is room for improvement in the decision-making of coordination triggering and cancellation.

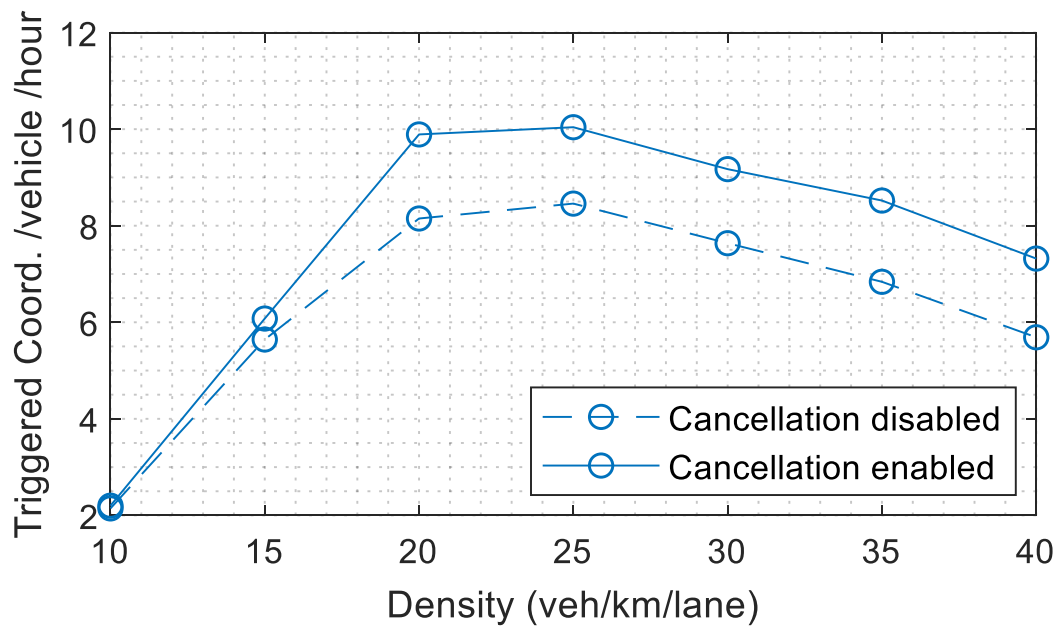

Fig. 4. Number of triggered coordinations per vehicle per hour.

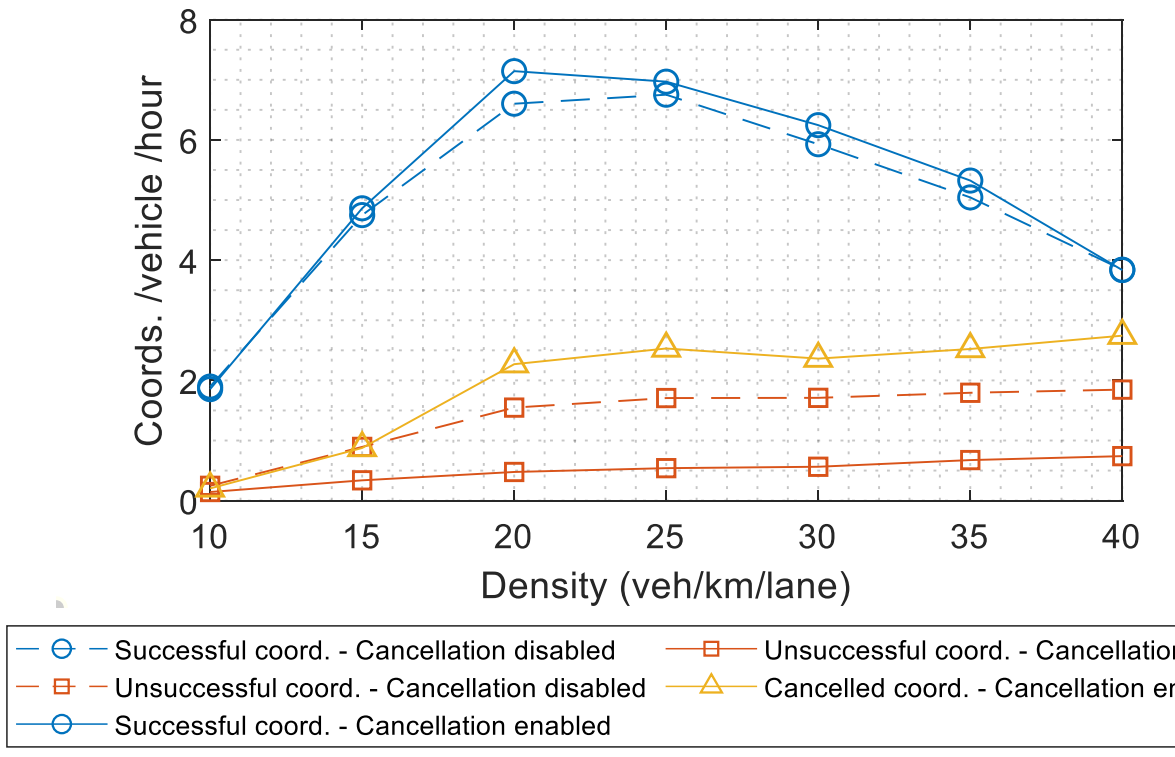

Fig. 5. Number of successful, unsuccessful, and cancelled coordinations per vehicle per hour with cancellation disabled and enabled.

## VI. CONCLUSIONS

This paper presented the design, implementation, and evaluation of maneuver coordination cancellation for connected automated driving. While existing research and standardization efforts have addressed how to initiate and execute coordinated maneuvers, our work focused on the complementary question of when and how to cancel coordinations that are not expected to end successfully. We proposed a full framework including a formal state machine, communication protocol, and initial decision-making mechanism, and assessed its performance in system-level simulations under varying traffic densities. The results showed that cancellation shortens the duration of unsuccessful coordinations, reducing wasted time and creating more opportunities for new maneuvers. This leads to a notable increase in the number of triggered coordinations and a measurable improvement in successful ones. At the same time, the findings revealed that many of the additional opportunities generated through cancellation also end in failure or cancellation, indicating that more advanced decision-making may help to fully exploit the potential of the cancellation mechanism. Overall, the study establishes maneuver coordination cancellation as an important building block for cooperative driving systems. It improves efficiency by reducing time spent on unsuitable interactions and provides a basis for future refinements.

## ACKNOWLEDGEMENTS

This work has been partially funded by MICIU/AEI/10.13039/501100011033 and "ERFD/EU" (PID2023-150308OB-I00), and the "European Union NextGenerationEU/PRTR" (TED2021-130436B-I00), as well as Generalitat Valenciana (CIAICO/2024/167).